\documentclass{article}
\usepackage{spconf,amsmath,graphicx,hyperref}
\usepackage{cite}
\usepackage{graphicx}
\usepackage{multirow}
\usepackage{booktabs}
\usepackage{makecell}
\usepackage{amsmath}

\title{LogPTR: Variable-Aware Log Parsing with Pointer Network}

\name{Yifan Wu$^{1*}$\thanks{$*$Work done during internship at Ant Group.}, Bingxu Chai$^{2}$, Siyu Yu$^{1}$, Ying Li$^{1\dagger}$\thanks{$^\dagger$Corresponding author. Email: li.ying@pku.edu.cn}, Pinjia He$^{3}$, Wei Jiang$^{2}$, Jianguo Li$^{2}$
\thanks{This work was supported by the China Postdoctoral Science Foundation under Grant Number 2025M781445, the Postdoctoral Fellowship Program of CPSF under Grant Number GZC20251085, and Ant Group through Ant Research Program.}
}

\address{$^{1}$ Peking University, Beijing, China \\ $^{2}$ Ant Group, Hangzhou, China \\ $^{3}$ The Chinese University of Hong Kong, Shenzhen, China}

\begin{document}
\ninept
\maketitle
\begin{abstract}
Due to the sheer size of software logs, developers rely on automated log analysis. Log parsing, which parses semi-structured logs into a structured format, is a prerequisite of automated log analysis. However, existing log parsers are unsatisfactory when applied in practice because they 1) ignore categories of variables, and 2) need labor-intensive model tuning.
To address these limitations, we propose LogPTR, a variable-aware log parser that can extract the static and dynamic parts in logs, and further identify categories of variables. 
The key of LogPTR is formulating log parsing as a text summarization problem and using a pointer mechanism to copy words from the log message and label tokens indicating categories of variables.
The experimental results on widely-used benchmark datasets show that LogPTR outperforms state-of-the-art log parsers on both general log parsing that extracts log templates and variable-aware log parsing that further identifies categories of variables.
\end{abstract}
\section{Introduction}
\label{sec:intro}
Logs play an important role in software systems to record system execution behaviors. By inspecting system logs, engineers can detect anomalies \cite{du2017deeplog}, localize software bugs \cite{chen2021pathidea}, or troubleshoot problems \cite{jia2021logflash} in the system.
However, facing the rapid growth in volume of logs, it is becoming more and more challenging to identify valuable information from enormous log data, even for experienced engineers.
To tackle this problem, \emph{automated log analysis} has emerged in recent years, aiming to automatically analyze log data \cite{he2021survey}.

\begin{figure}
    \centering
    \includegraphics[width=0.90\columnwidth]{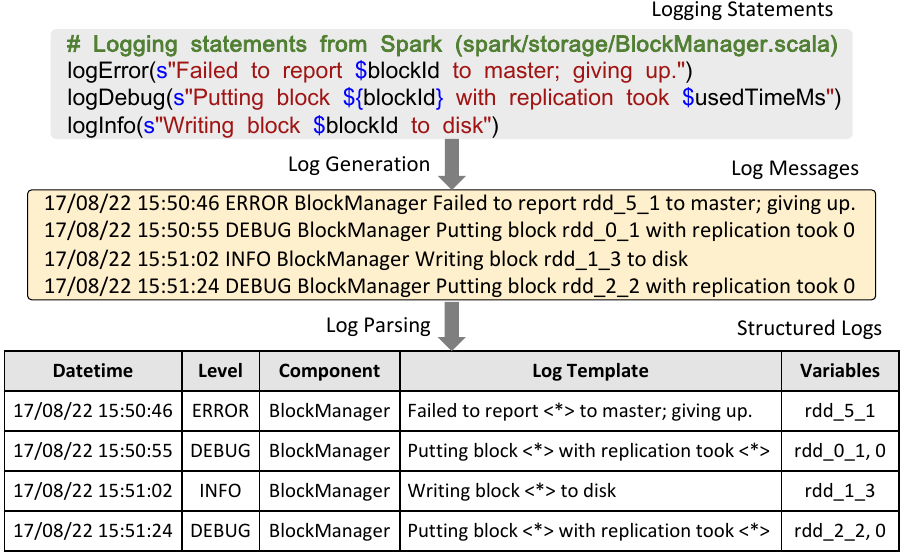}
    \caption{An example of log parsing from Spark.}
    \label{fig:example}
    \vspace{-2ex}
\end{figure}

The fundamental step for automated log analysis is \emph{log parsing}, which parses the \emph{semi-structured} log messages into a \emph{structured} format \cite{zhang2023system}.
As shown in \figurename~\ref{fig:example}, log messages are generated from logging statements in the source code.
A log message usually contains the log header (e.g., time, levels) and the message content, which further consists of two parts: 1) \emph{static} log template describing system events; 2) \emph{dynamic} variables, which vary during runtime and reflect system runtime status.
The goal of log parsing is to extract static log templates and dynamic variables from log messages.

\begin{figure*}
    \centering
    \includegraphics[width=0.85\textwidth]{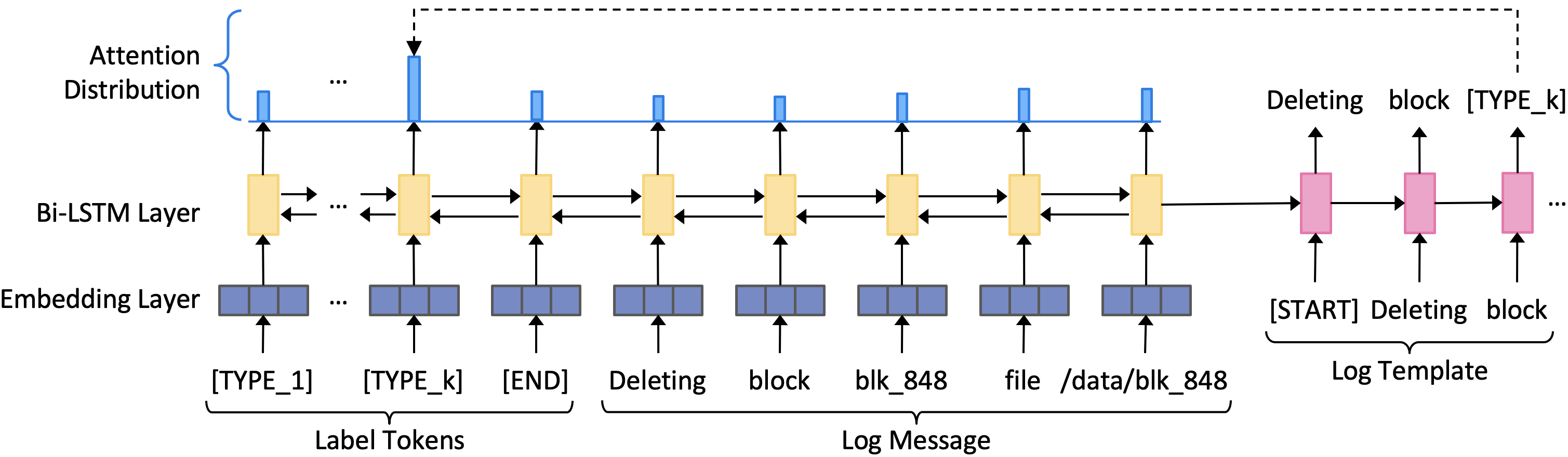}
    \caption{The model architecture of LogPTR.}
    \label{fig:model}
    \vspace{-2ex}
\end{figure*}

To achieve automated log parsing, many approaches have been proposed \cite{zhang2023system}, which can be categorized into syntax-based and semantic-based log parsers. 
Syntax-based log parsers \cite{fu2009execution,he2017drain,dai2020logram,nagappan2010abstracting}
utilize specific features or heuristics (e.g., log length and frequency) to extract the constant parts of log messages as templates. 
In contrast, semantic-based log parsers \cite{liu2022uniparser, le2023log, li2023did} employ deep learning models to learn
semantics from log data to parse log messages.

Although making progress, existing log parsers are still unsatisfactory when applied in practice. We have identified the following limitations of the existing log parsers:
\begin{itemize}
    \item \textbf{Ignoring categories of variables.} The categories of variables record valuable information to assist log understanding and analysis \cite{li2023did}. However, almost all log parsers aim to extract all dynamic variables of logs and output the remaining static words, which ignores the categories of variables.
    \item \textbf{Requiring labor-intensive model tuning.} Most log parsers require handcrafted rules (e.g., regular expressions) for data pre-processing and adjust hyperparameters (e.g., the number of clusters or similarity threshold) for different log datasets. When applying to a new log dataset, due to different logging formats and behaviors, the time-consuming adjustment of handcrafted rules and hyperparameters makes it suffer from system migrations.
\end{itemize}

To tackle the aforementioned limitations, we formulate log parsing as a text summarization problem and propose LogPTR, a simple yet effective variable-aware log parser with pointer network.
LogPTR can identify which words are constants, which words are variables, and what their corresponding categories.
Specifically, we first add label tokens at the beginning of log messages to indicate the variable categories.
Then, LogPTR generates the variable-aware log template by copying words from the log message and label tokens with the pointer mechanism.
LogPTR eliminates the need for handcrafted rules and applies a consistent set of hyperparameters across all datasets, thereby avoiding labor-intensive model tuning.

We evaluate LogPTR on widely-used benchmarks \cite{zhu2023loghub}.
LogPTR achieves a group accuracy of 0.990 and a parsing accuracy of 0.973 on average for general log parsing, which outperforms state-of-the-art baselines.
In addition, LogPTR effectively identifies variable categories for variable-aware log parsing and achieves an average parsing accuracy of 0.972, which is higher than baselines.
Furthermore, LogPTR shows robustness across different log datasets.

To summarize, our main contributions are as follows:
\begin{itemize}
    \item \textbf{Dimension}. This paper opens a new dimension that formulates log parsing as a text summarization problem. 
    By adding label tokens for variable categories, our approach can be easily extended to variable-aware log parsing.
    \item \textbf{Technique}. We implement LogPTR, a simple yet effective variable-aware log parser that can identify log templates, variables, and categories of variables.
    LogPTR does not require any handcrafted rules and uses the same set of hyperparameters across all datasets.
    \item \textbf{Evaluation}. 
    Extensive experiments on widely-used benchmark datasets demonstrate that LogPTR outperforms prior state-of-the-art approaches both in general log parsing and variable-aware log parsing.
\end{itemize}

\section{Methodology}
\label{sec:method}
\subsection{Problem Definition}
Intuitively, log messages can be regarded as a special type of natural language due to their semi-structured essence of mixing unstructured natural language and structured variables.
Thus, we formulate log parsing as a text summarization problem, where the output vocabulary consists of the tokens in the log message and the label tokens indicating categories of variables.
Specifically, for \emph{general log parsing}, i.e., extract log templates and variables in logs, we use one label token to indicate the token is a variable, regardless of its categories. While for \emph{variable-aware log parsing}, i.e., further identifying categories of variables, we add several label tokens corresponding to the number of categories of variables.

Formally, suppose there are \emph{m} categories of variables, denoted as $C=\{c_1,...,c_m\}$ and a raw log message consists of \emph{n} tokens after tokenization, denoted as $T=\{t_1,..,t_n\}$.
We take $X=[C,T]$ as input and generates the output $Y=\{y_1,...,y_k\} \ y_i \in [1, m+n]$, where $y_i$ indicates the index of the input.
Based on the indices, we can output the variable-aware log template $L=\{X{y_1},...,X{y_k}\}$ as the final result.

\begin{table*}
\centering
\caption{Accuracy comparison between LogPTR and baselines on general log parsing (\%).}
\label{tab:result_nva}
\begin{tabular}{c|cc|cc|cc|cc|cc|cc|cc|cc}
\toprule
\multirow{2}{*}{Dataset} & \multicolumn{2}{c|}{AEL} & \multicolumn{2}{c|}{Logram} & \multicolumn{2}{c|}{LenMa} & \multicolumn{2}{c|}{Drain} & \multicolumn{2}{c|}{UniParser} & \multicolumn{2}{c|}{LogPPT} & \multicolumn{2}{c|}{LILAC}     & \multicolumn{2}{c}{LogPTR}    \\ 
                         & GA             & PA     & GA           & PA           & GA              & PA      & GA              & PA      & GA            & PA            & GA               & PA      & GA            & PA            & GA            & PA            \\ \midrule
Hadoop                   & 86.9           & 26.2   & 45.1         & 11.3         & 88.5            & 8.3     & 94.8            & 26.9    & 94.8          & 59.0          & 80.1             & 54.2    & 98.3          & 63.2          & \textbf{99.5} & \textbf{96.5} \\
HDFS                     & 99.8           & 35.4   & 93.0         & 0.4          & 99.8            & 1.0     & 99.8            & 35.4    & 99.8          & \textbf{99.7} & 84.5             & 38.9    & \textbf{100}  & 50.0          & 99.7          & 99.6          \\
OpenStack                & 75.7           & 1.8    & 32.6         & 0.0          & 73.2            & 1.8     & 73.3            & 1.8     & 73.3          & 2.2           & \textbf{100}     & 90.6    & \textbf{100}  & 46.8          & \textbf{100}  & \textbf{99.4} \\
Spark                    & 90.5           & 35.9   & 28.2         & 25.9         & 88.3            & 0.4     & 92.2            & 36.2    & 92.2          & 90.4          & 84.8             & 96.1    & 99.9          & 98.3          & \textbf{100}  & \textbf{99.2} \\
Zookeeper                & 92.1           & 49.6   & 72.4         & 47.3         & 84.1            & 45.2    & 96.7            & 49.7    & 96.7          & 94.2          & 98.5             & 98.8    & 98.9          & 66.8          & \textbf{100}  & \textbf{98.9} \\
BGL                      & 95.7           & 34.2   & 58.7         & 12.4         & 69.0            & 8.2     & 96.3            & 34.2    & 96.3          & 80.5          & 58.4             & 82.2    & \textbf{98.2} & \textbf{96.5} & 90.3          & 93.7          \\
HPC                      & 90.3           & 65.9   & 91.0         & 64.3         & 83.0            & 63.2    & 88.7            & 63.5    & 88.7          & 72.8          & 99.0             & 92.5    & 97.0          & \textbf{99.2} & \textbf{99.4} & 97.9          \\
Thunderbird              & 94.1           & 3.6    & 55.4         & 0.4          & 94.3            & 2.5     & 95.5            & 4.6     & 95.5          & 76.5          & 65.7             & 88.8    & 98.1          & \textbf{94.6} & \textbf{98.6} & 93.9          \\
Linux                    & 67.2           & 16.9   & 36.1         & 12.4         & 70.1            & 12.2    & 69.0            & 18.3    & 69.0          & 23.6          & 18.5             & 42.4    & 75.2          & 33.7          & \textbf{100}  & \textbf{96.8} \\
Mac                      & 76.3           & 16.6   & 56.8         & 16.9         & 69.8            & 12.5    & 78.6            & 21.7    & 78.6          & 36.6          & 71.0             & 51.0    & 78.7          & 54.4          & \textbf{98.2} & \textbf{89.7} \\
Apache                   & \textbf{100}   & 69.4   & 31.2         & 0.6          & \textbf{100}    & 0.0     & \textbf{100}    & 69.4    & \textbf{100}  & 69.4          & 58.2             & 99.2    & \textbf{100}  & \textbf{100}  & \textbf{100}  & \textbf{100}  \\
OpenSSH                  & 53.6           & 24.5   & 61.1         & 29.8         & 92.6            & 13.3    & 78.9            & 50.8    & 78.9          & 46.6          & 43.6             & 97.5    & 75.3          & 80.6          & \textbf{99.7} & \textbf{98.9} \\
HealthApp                & 56.8           & 16.3   & 26.7         & 11.2         & 17.4            & 12.9    & 78.0            & 23.1    & 78.0          & 60.1          & \textbf{100}     & 66.8    & 91.7          & 77.0          & \textbf{100}  & 97.8          \\
Proxifier                & 49.5           & 0.0    & 50.3         & 0.0          & 50.8            & 0.0     & 52.6            & 0.0     & 52.6          & 50.3          & \textbf{100}     & 0.0     & \textbf{100}  & 47.3          & \textbf{100}  & \textbf{100}  \\ \midrule
Average                  & 80.6           & 28.3   & 52.8         & 16.6         & 77.2            & 13.0    & 85.3            & 31.1    & 85.3          & 61.6          & 75.9             & 71.4    & 93.7          & 72.0          & \textbf{99.0} & \textbf{97.3} \\ \bottomrule
\end{tabular}
\vspace{-2ex}
\end{table*}

\subsection{Model Architecture}
LogPTR adopts a sequence-to-sequence architecture.
The encoder consists of an embedding layer and a bidirectional LSTM (Bi-LSTM) layer, while the decoder consists of a unidirectional LSTM with the same hidden size as the encoder, and a pointer mechanism to copy words from the log message and label tokens indicating variable categories. The architecture of LogPTR is shown in \figurename~\ref{fig:model}.

\textbf{Embedding Layer.}
Engineers can create an infinite variety of variable names or abbreviations, which are far beyond the scale of common words. 
In addition, many variables in logs are digits (e.g., ``23869471'') or special symbols (e.g., ``/etc/data/''), which can have almost unlimited potential of creating ``new words'' based on different system behaviors.
This may incur the out-of-vocabulary (OOV) problem when encoding logs \cite{liu2022uniparser, li2023did}.
To solve this problem, we use WordPiece \cite{wu2016google} to tokenize the log message into a sequence of subwords.
Then, the embedding layer encodes each subword into a high-dimensional embedding vector as the next layer input.

\textbf{Bi-LSTM Layer.}
Similar to sentences in natural language, the log is a series of words that have sequential dependencies between the words. Hence, we use a Bi-LSTM to capture the dependencies between words in logs.
Specifically, the subword vector is fed into the Bi-LSTM layer. Afterward, the hidden state $e_i$ of Bi-LSTM is used for the pointer mechanism to generate words.

\textbf{Pointer Mechanism.}
We use the pointer mechanism \cite{vinyals2015pointer} to generate the variable-aware log template by creating pointers to words in the input sequence.
Specifically, on each step $t$, the decoder receives the word embedding of the previous word and has decoder hidden state $d_t$.
We obtain the following output distribution over the vocabulary of inputs:
\begin{align}
    u^t_i &= v^Ttanh(W_1e_i + W_2d_t) \\
    P(y_t|y_{<t},X) &= softmax(u^t)
\end{align}
where $v$, $W_1$, and $W_2$ are learnable parameters.

During training, LogPTR is trained with the maximum likelihood objective, defined as:
\begin{align}
    \mathcal{L} = -{\textstyle \sum_{t=0}^{T}} logP(y_t|y_{<t},X)
\end{align}

\section{EXPERIMENTAL DESIGN}
\label{sec:experiment}

\subsection{Datasets}
Our experiments for evaluating general log parsing are conducted on 14 widely-used benchmark datasets from LogHub \cite{zhu2023loghub}.
Furthermore, Li et al. \cite{li2023did} manually annotated the log datasets from LogHub with ten categories of variables. We use this dataset to evaluate variable-aware log parsing.
For each dataset, we randomly split it into training (20\%), validation (20\%), and testing data sets (60\%). 

\subsection{Baselines}
For general log parsing, we select seven open-source and state-of-the-art log parsers for comparison. These log parsers cover most categories of parsing techniques and achieve good results on public datasets \cite{khan2022guidelines}.
The first four are syntax-based log parsers, including frequency-based (i.e., AEL \cite{jiang2008abstracting} and Logram \cite{dai2020logram}), clustering-based (i.e., LenMa \cite{shima2016length}), and heuristic-based (i.e., Drain \cite{he2017drain}).
We also choose three semantic-based log parsers, including tuning-based (i.e., UniParser \cite{liu2022uniparser} and LogPPT \cite{le2023log}) and prompt-based (i.e., LILAC \cite{jiang2023lilac}).
To ensure a fair comparison, we use the implementations of all baselines from their replication repositories, choosing the default settings or hyperparameters.
For variable-aware log parsing, we use VALB \cite{li2023did} for comparison, which is the only baseline that can identify variable categories. Since the code of VALB is not open-source, we utilize the reported results for comparison.

\subsection{Evaluation Metrics}
Following recent studies \cite{liu2022uniparser, le2023log, khan2022guidelines, zhu2019tools}, we apply two widely-used metrics in our evaluation, including:

\textbf{Group Accuracy (GA).}
Group Accuracy \cite{zhu2019tools} is the most commonly used metric for log parsing.
GA considers log parsing as a clustering process where log messages with different log templates are clustered into different groups \cite{khan2022guidelines}.
GA is defined as the ratio of “correctly parsed” log messages over the total number of log messages, where a log message is considered “correctly parsed” if and only if it is grouped with other log messages consistent with the ground truth.

\textbf{Parsing Accuracy (PA).}
Parsing Accuracy (or Message-Level Accuracy \cite{liu2022uniparser}) is defined as the ratio of “correctly parsed” log messages over the total number of log messages,
where a log message is considered to be “correctly parsed” if and only if every token of the log message is correctly identified as a template or variable in general log parsing, or further all the categories of variables are also correctly identified in variable-aware log parsing.

\subsection{Implementation Details}
We conducted experiments on a GPU Server equipped with an NVIDIA Tesla V100 GPU.
The embedding layer was configured with an embedding size of 256, and the LSTM layer had a hidden size of 256.
To mitigate overfitting, we applied a dropout rate of 0.2 to both embedding and LSTM layers.
We used the Adam optimizer with an initial learning rate of 0.001.
During the training phase, we set the batch size to 32 and trained the model for 100 epochs.

\section{EXPERIMENTAL RESULTS}

\subsection{Effectiveness on General Log Parsing}

\tablename~\ref{tab:result_nva} presents the accuracy of LogPTR and baselines on general log parsing. From the results, we can see that LogPTR outperforms baselines on almost all datasets in the two metrics.
Specifically, in terms of GA, LogPTR achieves the highest average accuracy of 99\% and the best results on 12 out of 14 datasets. It is worth noting that LogPTR achieves an accuracy of over 98\% on 13 datasets and achieves 100\% accuracy on 7 datasets among them, which is significantly superior to existing log parsers.
In terms of PA, LogPTR also achieves the highest average accuracy of 97.3\%.
Overall, the experimental results confirm that LogPTR is effective in general log parsing, which groups logs into the same templates and identifies correct templates and variables.

\begin{figure}
    \centering
    \includegraphics[width=0.99\columnwidth]{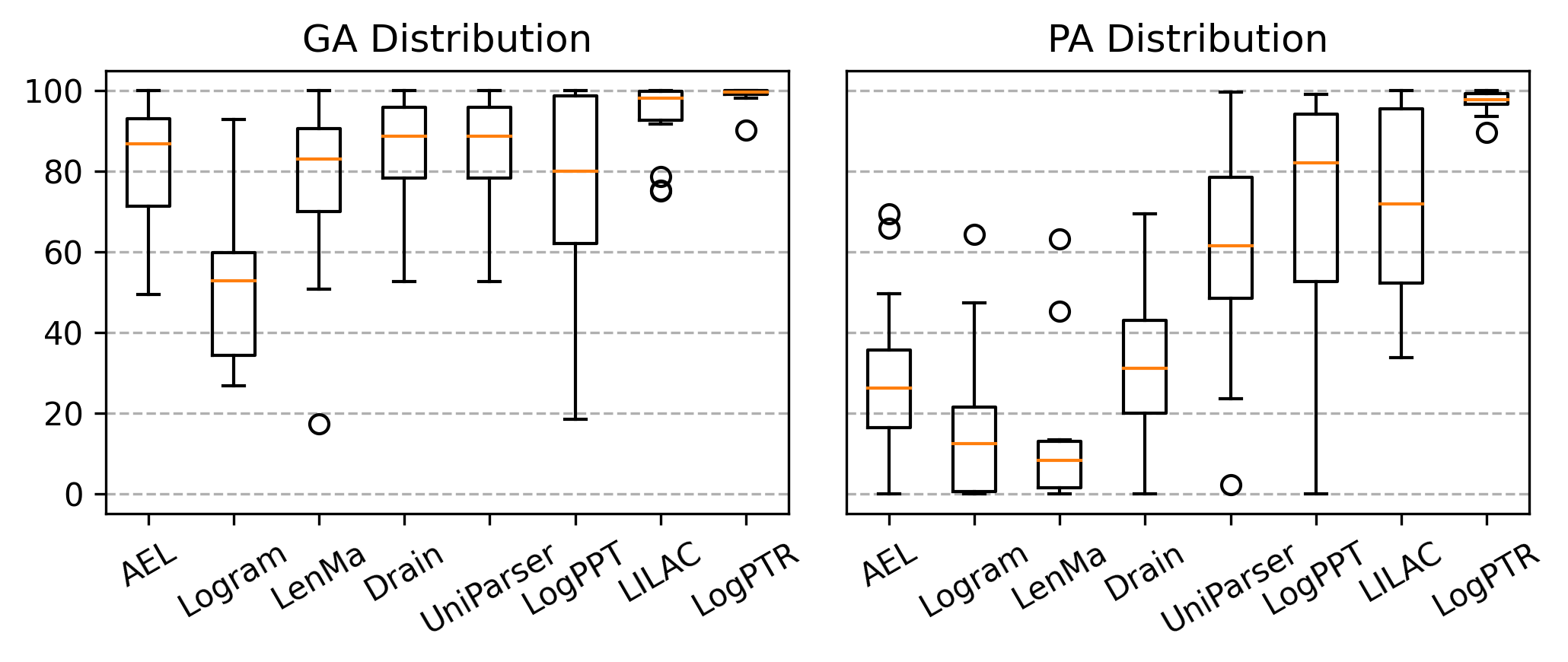}
    \caption{Robustness comparison between LogPTR and baselines.}
    \label{fig:robust}
    \vspace{-2ex}
\end{figure}

LogPTR explicitly aims at supporting a broad range of diverse log datasets because robustness is crucial to the use of a log parser in practice \cite{zhu2019tools}.
Therefore, we analyze and compare the robustness of LogPTR and baselines against different types of logs.
The results are shown in \figurename~\ref{fig:robust}. From the results, we can see that LogPTR outperforms baselines in terms of robustness across different log types.
Existing methods require different regular expressions for pre-processing and different hyper-parameter values, which involve domain-specific knowledge and labor-intensive model tuning. Thus, they perform inconsistently on different datasets.
In contrast, LogPTR does not require manually defining regular expressions and achieves the smallest variance over different datasets.
Moreover, LogPTR uses the same set of hyperparameter values in the training phase and does not require re-adjustment for each dataset.
Overall, the experimental results confirm that LogPTR is robust and can be applied to different log datasets with low effort.
 
\subsection{Effectiveness on Variable-aware Log Parsing}
\tablename~\ref{tab:result_va} presents the accuracy of LogPTR and VALB on variable-aware log parsing. From the results, we can see that LogPTR achieves a high parsing accuracy ranging from 86.6\% in Mac to 100\% in Apache, which is also close to the parsing accuracy as discussed in RQ1.
The average parsing accuracy of LogPTR is 97.2\%, which is higher than the baseline.
Moreover, LogPTR achieves the best results on 14 out of 16 datasets.
As for robustness, we calculate the standard deviation (STD) of parsing accuracy, and LogPTR achieves a smaller STD than the baseline (0.036 versus 0.045).
Overall, the experimental results confirm that apart from general log parsing, LogPTR can also efficiently identify the categories of variables in logs to perform variable-aware log parsing.

\begin{table}
\centering
\caption{Accuracy comparison between LogPTR and baselines on variable-aware log parsing (\%).}
\label{tab:result_va}
\begin{tabular}{p{2cm}<{\centering}|p{1.5cm}<{\centering}|p{1.5cm}<{\centering}}
\toprule
Dataset & VALB & LogPTR \\ \midrule
Android & 91.6 & \textbf{94.3} \\
Apache & 99.3 & \textbf{100} \\
BGL & 89.6 & \textbf{95.9} \\
Hadoop & 96.8 & \textbf{96.8} \\
HDFS & 96.5 & \textbf{99.8} \\
HealthApp & \textbf{98.8} & 97.1 \\
HPC & 99.0 & \textbf{99.2} \\
Linux & 95.9 & \textbf{96.1} \\
Mac & 86.2 & \textbf{86.6} \\
OpenSSH & 97.6 & \textbf{99.4} \\
OpenStack & 93.2 & \textbf{99.8} \\
Proxifier & 100 & \textbf{100} \\
Spark & 99.1 & \textbf{99.7} \\
Thunderbird & 87.8 & \textbf{92.5} \\
Windows & \textbf{99.0} & 98.3 \\
Zookeeper & 98.1 & \textbf{98.9} \\ \midrule
Average & 95.5 & \textbf{97.2} \\ \bottomrule
\end{tabular}
\vspace{-2ex}
\end{table}

\section{RELATED WORK}
\label{sec:relatedwork}
Existing Log parsers can be categorized into syntax-based and semantic-based log parsers.
Syntax-based log parsers leverage manually designed features to extract log templates, which have been widely explored in the past. Specifically, syntax-based log parsers can be further subdivided into three categories.
(1) \textit{Frequency-based} parsers: These log parsers \cite{vaarandi2003data,nagappan2010abstracting,vaarandi2015logcluster,dai2020logram} utilize frequent patterns of token position or n-gram information to distinguish the templates and variables in log messages.
(2) \textit{Clustering-based} parsers: These log parsers \cite{fu2009execution,tang2011logsig,shima2016length,hamooni2016logmine} compute similarities between log messages to cluster them into different groups and then extract the constant parts of log messages.
(3) \textit{Heuristic-based} parsers: These log parsers \cite{du2016spell,he2017drain,jiang2008abstracting,makanju2009clustering,yu2023brain,liu2024xdrain} encode expert domain knowledge into general and effective heuristic rules to identify log templates. 
Semantic-based log parsers are a recently emerging class of parsers and can achieve higher parsing accuracy by mining semantics from log messages. Specifically, Semantic-based log parsers can be further subdivided into two categories.
(1) \textit{tuning-based} parsers: These log parsers \cite{liu2022uniparser,li2023did,le2023log,yu2023self,yu2023log} typically need labeled log data for model training or tuning. For example, Uniparser \cite{liu2022uniparser} formulates log parsing as a token classification problem, employing Bi-LSTM models for training.
VALB \cite{li2023did} formulates log parsing as a sequence tagging problem and uses Bi-LSTM-CRF models for training.
(2) \textit{prompt-based} parsers: These log parsers \cite{wu2025log,jiang2023lilac,hong2025cslparser} utilize a designed prompt to query language language models (LLMs) for log parsing.
Unlike syntax-based log parsers, LogPTR doesn't require any handcrafted features or heuristic rules, which can adapt to new log datasets with low effort. Furthermore, compared to semantic-based log parsers, LogPTR can not only identify log templates and variables but also identify the categories of variables, which can aid log understanding and analysis.

\section{CONCLUSION}
This paper proposes LogPTR, a simple yet effective variable-aware log parser that can extract static and dynamic parts in logs and further identify categories of variables.
The key of LogPTR is formulating log parsing as a text summarization problem and using a pointer mechanism to copy words from the log message and label tokens indicating categories of variables.
Experimental results show that LogPTR outperforms the state-of-the-art log parsers on both general log parsing and variable-aware log parsing.

\bibliographystyle{IEEEbib}
\bibliography{mybibfile}

\end{document}